\documentclass{emulateapj}

\slugcomment{To Appear in the Astrophysical Journal Letters}

\shorttitle{Large-Scale Structures with Forming Clusters at Redshift 6}
\shortauthors{Ouchi et al.}

\begin{document}

\title{The Discovery of Primeval Large-Scale Structures\\
       with Forming Clusters at Redshift 6\altaffilmark{1}}

\author{Masami Ouchi        \altaffilmark{2,3},
        Kazuhiro Shimasaku  \altaffilmark{4},
	Masayuki Akiyama    \altaffilmark{5},
	Kazuhiro Sekiguchi  \altaffilmark{5},\\
	Hisanori Furusawa   \altaffilmark{5},
	Sadanori Okamura    \altaffilmark{4},
	Nobunari Kashikawa  \altaffilmark{6},
	Masanori Iye        \altaffilmark{6},\\
	Tadayuki Kodama     \altaffilmark{6},
	Tomoki Saito        \altaffilmark{4},
	Toshiyuki Sasaki    \altaffilmark{5},
	Chris Simpson       \altaffilmark{7},\\
	Tadafumi Takata     \altaffilmark{5},
	Toru Yamada         \altaffilmark{6},
	Hitomi Yamanoi      \altaffilmark{8},\\
	Makiko Yoshida      \altaffilmark{4}, and
	Michitoshi Yoshida  \altaffilmark{9}
	}

\altaffiltext{1}{Based on data collected at 
        Subaru Telescope, which is operated by 
        the National Astronomical Observatory of Japan.}
\altaffiltext{2}{Space Telescope Science Institute,
        3700 San Martin Drive, Baltimore, MD 21218, USA; ouchi@stsci.edu.}
\altaffiltext{3}{Hubble Fellow}
\altaffiltext{4}{Department of Astronomy, School of Science,
        University of Tokyo, Tokyo 113-0033, Japan}
\altaffiltext{5}{Subaru Telescope, National Astronomical Observatory, 
        650 N.A'ohoku Place, Hilo, HI 96720, USA}
\altaffiltext{6}{National Astronomical Observatory, 
        Tokyo 181-8588, Japan}
\altaffiltext{7}{Department of Physics, University of Durham,
        South Road, Durham DH1 3LE, UK}
\altaffiltext{8}{Department of Mathematical and Physical Sciences,
        Faculty of Science, Japan Women's University, Tokyo 112-8681, Japan}
\altaffiltext{9}{Okayama Astrophysical Observatory,
    National Astronomical Observatory, Kamogata, Okayama 719-0232, Japan}

\begin{abstract}
We report the discovery of primeval large-scale structures (LSSs)
including two proto-clusters in a forming
phase at $z=5.7$. We carried out extensive deep narrow-band imaging
in the 1 deg$^2$ sky of the Subaru/XMM-Newton Deep Field, and
obtained a cosmic map of 515 Ly$\alpha$ emitters (LAEs) 
in a volume with a transverse dimension of
180 Mpc $\times$ 180 Mpc and a depth of 
$\sim 40$ Mpc in comoving units.
This cosmic map shows filamentary LSSs, including
clusters and surrounding 10-40 Mpc scale voids,
similar to the present-day LSSs. 
Our spectroscopic follow-up observations
identify overdense regions in which two dense clumps of LAEs 
with a sphere of $1$-Mpc diameter in physical units are included.
These clumps show about 130 times higher star-formation rate
density, mainly due to a large overdensity, $\sim 80$, of LAEs.
These clumps would be clusters in a formation phase 
involving a burst of galaxy formation.
\end{abstract}


\section{Introduction}
Recently, there has been a great progress in observational 
studies of structures of the high-redshift Universe.
\citet{steidel1998} found a proto-cluster at $z=3.1$
with a large population of Lyman break galaxies (LBGs).
A significant excess of Ly$\alpha$ emitters (LAEs)
was found by \citet{venemans2002}
around a radio galaxy at $z=4.1$, and a subsequent study
concluded that this excess is indeed a proto-cluster
with a size of $3-5$ Mpc \citep{miley2004}. 
More recently, \citet{venemans2004}
reported clustering of 6 LAEs
near a radio galaxy at $z\sim 5.2$.
For structures beyond cluster scales,
wide-field narrow-band surveys 
revealed very inhomogeneous distribution of LAEs
\citep{ouchi2003,shimasaku2003,
ajiki2003,shimasaku2004,hu2004} at $z\simeq 5-6$.
\citet{shimasaku2003} report 
an elongated overdense region of LAEs at $z=4.86$
on the sky of 20 Mpc in width and 50 Mpc in length.
However, on the same sky, \citet{shimasaku2004} found
no large-scale structure at $z=4.79$ which is closer to us
by 39 Mpc, indicating a large cosmic variance
over their surveyed volumes. Although 
a few segments of high-$z$ structures have been
found to date, the whole picture of high-$z$ Universe is veiled.
Moreover, we do not know when and how the structures of galaxies 
seen at present formed from the initial matter density 
fluctuations.
We started a systematic survey
to map the high-$z$ Universe with LAEs and LBGs at $3<z<7$ 
in the 1 deg$^2$ sky of the Subaru/XMM-Newton Deep Field 
(SXDF: R.A.$=2^h18^m00^s$, decl.$=-5^\circ 00 ' 00''$[J2000];
\citealt{sekiguchi2004}). We obtained cosmic
maps on scales of much larger than present-day LSSs 
to cover a network of LSSs at several redshifts 
between $z=3$ and $7$.
In this letter, we report the initial results
of our survey about structures made of LAEs at $z=5.7$.
Throughout this paper, we adopt
$H_0=70 h_{70} $km s$^{-1}$ Mpc$^{-1}$ and
$[\Omega_m,\Omega_\Lambda,n,\sigma_8]=[0.3,0.7,1.0,0.9]$.

\section{Photometric Sample}
We carried out extensive deep narrow-band imaging
on 2003 September 27-30 and October 22
with Subaru/Suprime-Cam \citep{miyazaki2002}.
We used the narrow-band filter $NB816$ whose
central wavelength and FWHM are $8150$\AA\ and 120\AA,
respectively \citep{hayashino2003,ajiki2003,hu2004}.
This filter enables us to
identify LAEs, galaxies with a strong 
Ly$\alpha$ emission line, at $z\simeq 5.7\pm 0.05$
with a small fraction of foreground contaminants (20-30\%;
e.g. \citealt{hu2004}).
Our 20-hour exposure imaging covered an area of 1.04 deg$^2$
with 5 Suprime-Cam pointings. 
The sky was clear and the observational conditions were stable, 
with typical seeing sizes of $0''.5-0''.6$.
The data were reduced with a software package SDFRED
\citep{yagi2002,ouchi2004a}.
The sky noise of the reduced image
is $26.0\pm 0.1$ (AB) 
at the $5\sigma$ level with a
$1''.8$-diameter circular aperture.

\begin{figure}
\epsscale{0.7}
\plotone{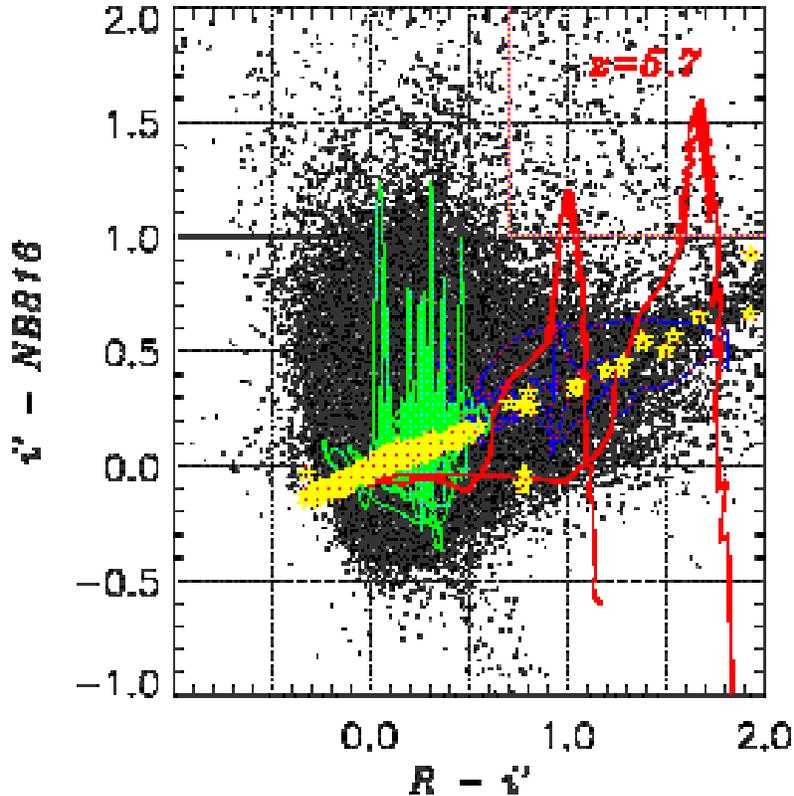}
\caption{
Two color diagrams for
continuum color ($R-i'$)
and narrow-band excess color ($i'-NB816$).
The black dots indicate colors of 
305,012 objects detected with $NB816<26.0$.
If the magnitude of an object is fainter than the $1\sigma$
magnitude ($1\sigma$ sky fluctuation), then the magnitude is
replaced with the $1\sigma$ magnitude.
All the curves show
colors of model galaxies at various redshifts.
Red lines indicate model LAEs which are
a composite spectrum of a 0.03 Gyr single burst
model galaxy (GISSEL00;\citealt{bruzual2003})
and a Ly$\alpha$ emission
($EW_0=22$\AA);
from the left to right, two different amplitudes
of inter-galactic medium absorption are applied:
$0.5\tau_{\rm eff}$ and $\tau_{\rm eff}$,
where $\tau_{\rm eff}$ is
\citeauthor{madau1995}'s \citeyearpar{madau1995}
median opacity.
The narrow-band excess in each of the peaks in
the red lines
indicates the Ly$\alpha$ emission
of LAEs at $z=5.7$.
Green lines show 6 templates of
nearby starburst galaxies \citep{kinney1996} up to $z=2$,
which are 6 classes of
starburst galaxies 
with $E(B-V)=0.0-0.7$.
The narrow-band excess
peaks in the green lines
correspond to the emission lines of
H$\alpha$ ($z=0.2$), [O{\sc iii}]($z=0.6$), H$\beta$($z=0.7$),
or [O{\sc ii}]($z=1.2$).
Blue lines show colors of typical
elliptical, spiral, and irregular galaxies \citep{coleman1980}
which are redshifted from $z=0$ to $z=3$. Yellow star marks show
175 Galactic stars given by \citet{gunn1983}.
The pink box surrounding the upper right region is the
selection criteria of our $z=5.7$ LAEs.
Because the distribution of the dots show a single track whose colors
are consistent with that of the Galactic stars, our photometry
is homogeneous over the 1 deg$^2$ sky of the SXDF.
{\bf (The paper with high resolution images can be
downloaded from http://www-int.stsci.edu/$\sim$ouchi/work/astroph/sxds\_LSS/)}
\label{fig:cc_NB816_modelobj}}
\end{figure}

We detected 305,012 objects down to $NB816=26.0$ 
with SExtractor \citep{bertin1996} and 
measured their narrow-band excess color ($i'-NB816$)
and continuum color ($R-i'$)
by combining existing very deep 
$R$- and $i'$-band images of the SXDF (see \citealt{kodama2004}).
Figure \ref{fig:cc_NB816_modelobj} shows the two color 
diagram of $i'-NB816$ and $R-i'$ for the $NB816$ detected
objects, together with colors of model galaxies and Galactic
stars. In Figure \ref{fig:cc_NB816_modelobj}, 
model LAEs at $z=5.7$ have very red colors both in 
$i'-NB816$ and $R-i'$
because of a strong Ly$\alpha$ emission line and a continuum 
trough at $<1216$\AA, which are distinguished from colors of 
foreground objects.
We defined the criteria for $z\simeq 5.7\pm 0.05$ LAEs
as $i'-NB816>1.0$ and $R-i'>0.7$, and selected
515 LAEs from all the detected objects.
These selection criteria isolate
emission line objects with an observed-frame
equivalent width of $EW_{\rm obs} \gtrsim 226$\AA\ and
a line flux of $f \gtrsim 6.1 \times 10^{-18}$ 
erg s$^{-1}$ cm$^{-2}$, if a flat continuum spectrum
is assumed.
The contamination rate of our LAE sample is estimated
to be about 30\% 
based on our spectroscopic follow-up observations
described in section 3.2.
The surface density and number density 
of these 515 LAEs
are estimated to be 
${\bar \Sigma}=0.14\pm0.01$ arcmin$^{-2}$ and 
${\bar n}=5.5 \pm 0.2 \times 10^{-4}$ Mpc$^{-3}$.
We also calculate the surface density down to the
same magnitude limit as \citet{hu2004} ($NB816<25.05$)
and obtain $0.03$ arcmin$^{-2}$ which is consistent
with that of \citet{hu2004}.

\section{Results and Discussion}
\subsection{Primeval Large-Scale Structures at $z=5.7$}
\begin{figure}
\epsscale{1.10}
\plotone{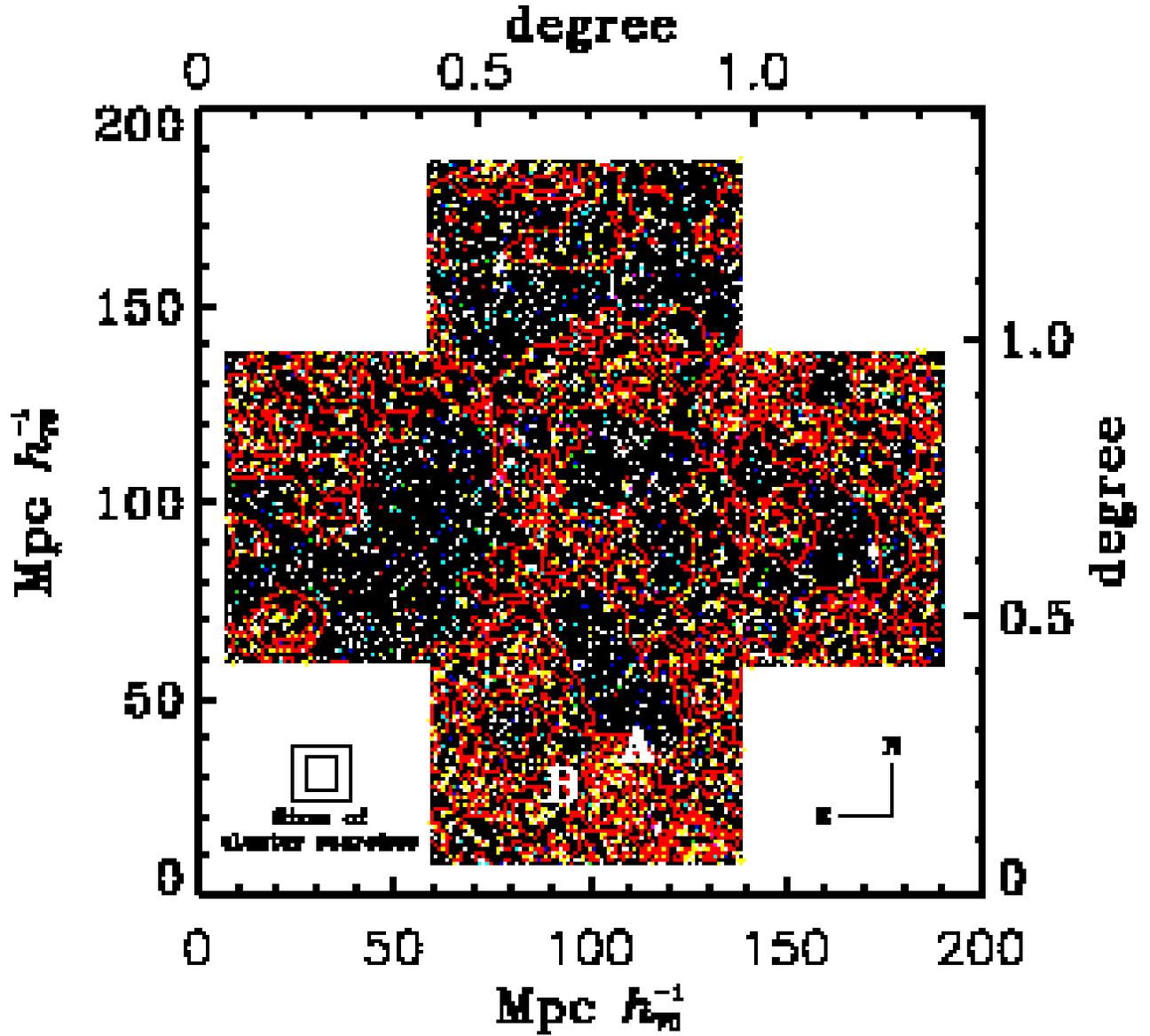}
\caption{
The distribution of $z=5.7 \pm 0.05$ LAEs
in the SXDF.
The positions of LAEs are shown with yellow dots.
The red lines correspond to contours of galaxy overdensity
from $\delta_\Sigma = -0.25$ to $3.25$
with a step of $\Delta=0.50$. The characters, A and B,
denote the positions of the two dense regions.
The scale on the map is marked in both degrees and (comoving) megaparsecs.
The large and small squares in the bottom left corner show the sizes of
the surveyed areas in the previous proto-cluster searches by
\citet{venemans2002} and \citet{miley2004}, respectively.
{\bf (The paper with high resolution images can be
downloaded from http://www-int.stsci.edu/$\sim$ouchi/work/astroph/sxds\_LSS/)}
\label{fig:density_cont_nb816_2}}
\end{figure}

Figure \ref{fig:density_cont_nb816_2} is the cosmic map 
showing the distribution of 515
LAEs at $z=5.7$ in the SXDF.
The surveyed volume has a transverse dimension of
180 Mpc $\times$ 180 Mpc and a depth of
$\sim 40$ Mpc in comoving units.
This is the first cosmic map, ever obtained,
covering a $>100$ Mpc square area of
the Universe at any high redshifts ($z>2$).
In Figure \ref{fig:density_cont_nb816_2}, 
the LAEs have a very clumpy distribution,
forming concentrations with a typical size of a few Mpc,
comparable in size to the proto-cluster found by
\citet{venemans2002,miley2004}.
These concentrations are not isolated, but connected with
one another by filamentary overdense regions.
The elongated overdense region of LAEs found  
by \citet{shimasaku2003} at $z=4.86$ may be a segment of
a descendant of such a filamentary structure.
There are also found several voids of ellipsoidal shapes
with sizes of 10-40 Mpc in which almost no galaxy exists.
The characteristic sizes of filaments and voids seen
in our map are comparable to those of the present-day Universe.
Thus, this map marks the discovery of primeval LSSs at $z=5.7$.

We quantify the large-scale clumpiness of the galaxy distribution
by estimating $\sigma_{20}$, the rms fluctuation of
galaxy overdensity within a sphere of 20 Mpc
(comoving) radius.
Considering that the radial depth of the surveyed volume
is about 40 Mpc, 
we estimate the fluctuation by
$\sigma_{20}^2 \simeq \sigma_{\Sigma20}^2 =
[\left<(\Sigma_{20}-\bar{\Sigma}_{20})^2\right>-\bar{\Sigma}_{20}]/
\bar{\Sigma}_{20}^2$,
where $\sigma_{\Sigma20}$ is the rms surface
overdensity within circles of
20 Mpc radius, and $\Sigma_{20}$ and $\bar{\Sigma}_{20}$ are 
the observed number
and the mean number of LAEs in a circle \citep{peebles1980}.
We obtain $\sigma_{20}=0.4\pm 0.2$.
This $\sigma_{20}$ is comparable to or 
at least a half the value for the present-day LSSs, i.e.
$\sigma_{20}(z=0)=0.5-0.6$, obtained from
galaxies with $b_j \leq 17.5$ \citep{seaborne1999}.
The characteristic shapes and the rms fluctuations indicate that
these primeval LSSs at $z\sim 6$ are 
similar to the present-day LSSs. 
We will present results of more detailed analyses 
for these primeval LSSs, such as counts-in-cell,
in our forthcoming paper.

\subsection{Two Clumps Identified by Spectroscopy: \\
Forming Clusters?}
\begin{figure}
\epsscale{1.10}
\plotone{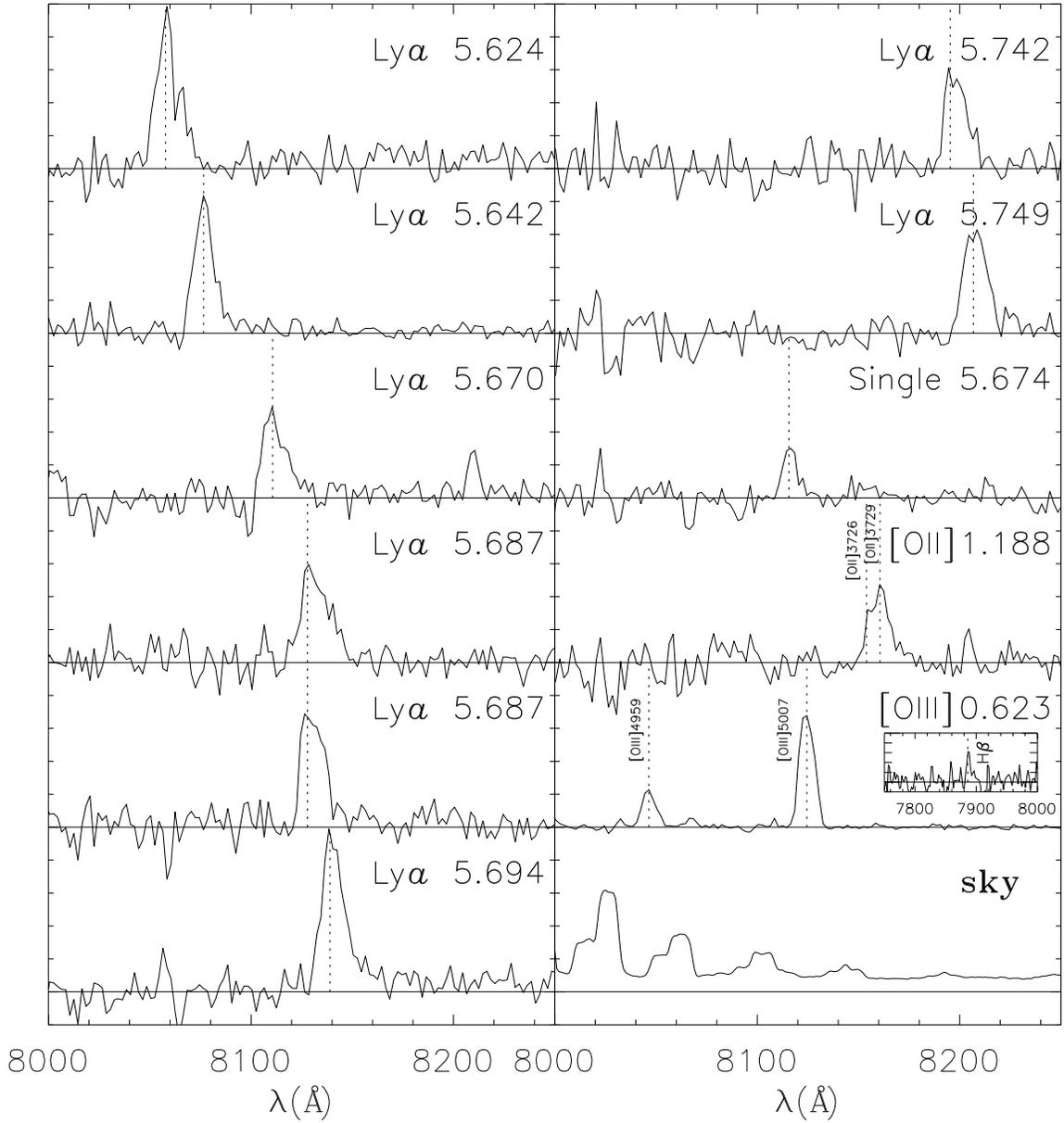}
\caption{
Spectra of eight secure LAEs,
together with an unresolved-single line
emitter and [O{\sc ii}] and [O{\sc iii}] emitters.
The caption on
each panel indicates the classification and redshift.
Ticks on the vertical axis denote $3\times 10^{-19}$
ergs s$^{-1}$ cm$^{-2}$ \AA$^{-1}$, except for the following
two panels because of display purpose.
The second-top left and the fifth-top right panels
have ticks of $6\times 10^{-19}$ and $1.5\times 10^{-18}$
ergs s$^{-1}$ cm$^{-2}$ \AA$^{-1}$, respectively.
The relative intensity of
the night-sky emission is shown on the bottom right.
\label{fig:spec_disp_paper}}
\end{figure}

We find that among the dense concentrations seen in
Figure \ref{fig:density_cont_nb816_2}, 
the one at $2^h 17^m 47.2^s$, $-5^\circ 28 ' 40''$[J2000]
has the highest density contrast, $\delta_\Sigma=3.3$,
with the $4.8\sigma$ significance level,
where $\delta_\Sigma$ is the surface overdensity 
for a circle of 8 Mpc radius.
We refer to this concentration as Region A.
Region B, a neighboring concentration at
$2^h 18^m 19.6^s$,$-5^\circ 32 ' 52''$[J2000],
also has a high density contrast of
$\delta_\Sigma=1.5$ ($2.2\sigma$).
In order to obtain three-dimensional distributions
of LAEs in Regions A and B,
we carried out spectroscopy for LAEs with
Subaru/FOCAS \citep{kashikawa2002}
on 2003 December 20, 23, and 25.
We placed a slit mask on each region.
Another mask was put on a control field well separated
from Regions A and B.
We made a 2-hour exposure for each mask, and obtained 22 spectra
in total, whose spectral range and resolution
are $\lambda$=4900-9400\AA\ and $\lambda/\Delta \lambda \simeq 1000$,
respectively.
Among the 22 objects, we classify one as an [O{\sc iii}] emitter
at $z=0.6$, two as [O{\sc ii}] emitters at $z=1.2$,
and eight as securely identified LAEs
with a characteristic asymmetric feature in Ly$\alpha$ line
whose skewness exceeds 1.5 with the $3\sigma$ level.
We show spectra of these emitters in Figure \ref{fig:spec_disp_paper}.
Since we have eight secure LAEs
among the 11 (=8+2+1) identified objects,
we estimate the contamination rate of our LAE
sample to be about 30\%, which is
consistent with those of other
LAE samples \citep{rhoads2003,shimasaku2003,hu2004}.
The other 11 objects out of the 22 are
emitters with an unresolved-single line
at $\sim 8150$\AA.
Since most unresolved-single line emitters
have been found to be LAEs with a moderate velocity dispersion
($\sim 200$km s$^{-1}$; \citealt{hu2004}),
we also regard these 11 objects as LAEs. Thus
we have 19 LAEs with spectroscopic redshifts
as summarized in Table \ref{tab:lae_spec}.

\begin{figure}
\epsscale{1.00}
\plotone{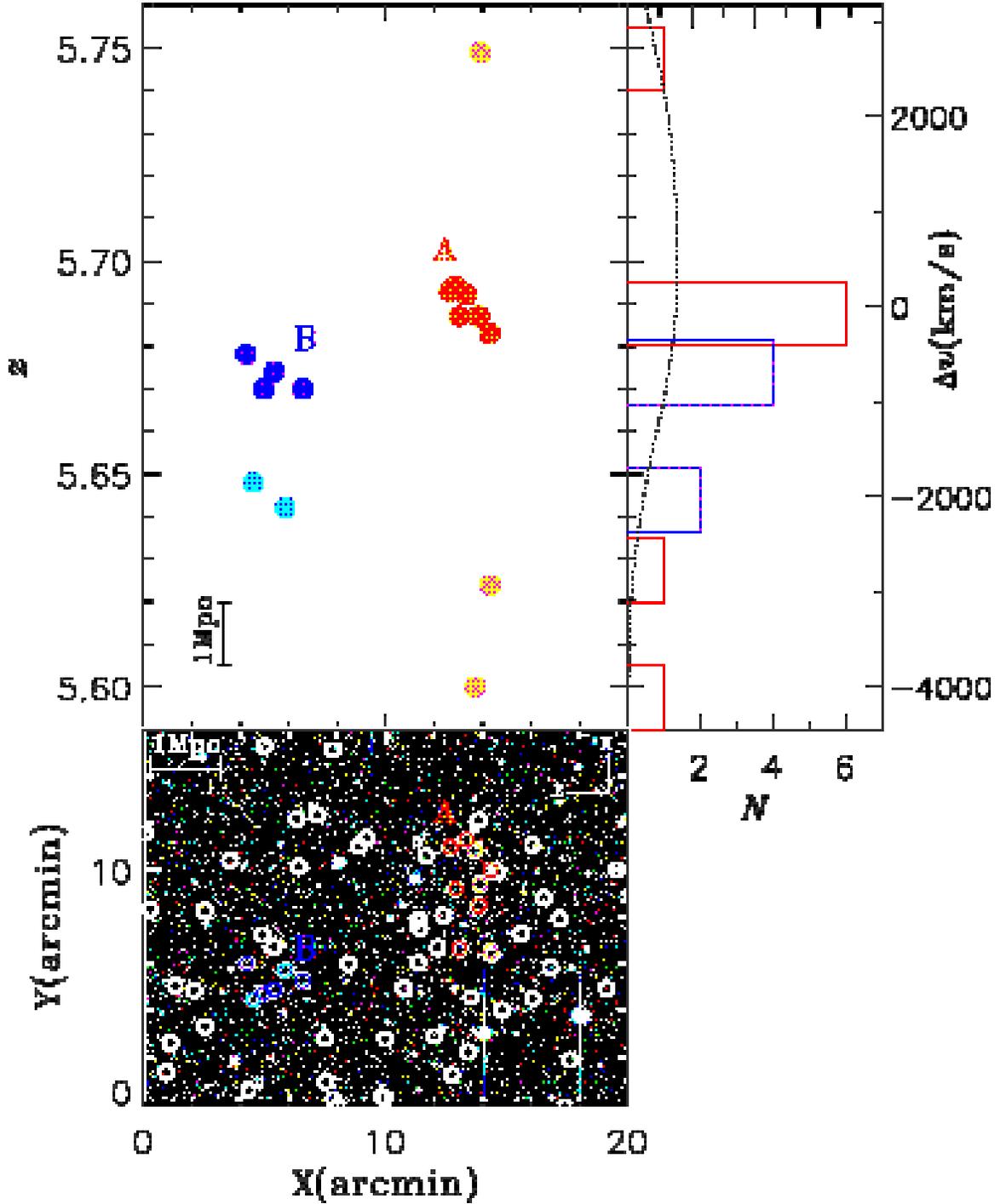}
\caption{
Three dimensional map of Regions A and B.
The upper left panel shows the distribution of LAEs
in transverse (East to the West) vs. radial (redshift) directions,
while the bottom panel represents the distribution of
LAEs projected on the sky. The red and blue
circles correspond to the LAEs
associated with the forming clusters (Clumps A and B), respectively,
while the orange and cyan circles denote other
LAEs in these regions.
The white open circles plotted on the bottom panel denote
the positions of LAEs without spectroscopic redshift.
The vertical and horizontal bars
indicate the length of 1 Mpc in physical units.
The upper right panel shows the redshift distribution of
LAEs with spectroscopic redshifts.
The red and blue histograms
correspond to LAEs in Regions A and B, respectively.
The dotted line indicates the mean-expected number of LAEs
for each region.
{\bf (The paper with high resolution images can be
downloaded from http://www-int.stsci.edu/$\sim$ouchi/work/astroph/sxds\_LSS/)}
\label{fig:threeDmap_sxdss_plushist}}
\end{figure}

We plot in Figure \ref{fig:threeDmap_sxdss_plushist}
the three-dimensional distribution of LAEs
for Regions A and B.
The histogram of Figure \ref{fig:threeDmap_sxdss_plushist} 
shows that these regions have
two clumps made of 10 LAEs with a significant excess.
Each of the clumps has a diameter of
about 1 Mpc in physical units (7 Mpc in comoving units).
For these clumps, the 1-dimensional velocity dispersion of galaxies
is fairly small,
$\sim 180$ km s$^{-1}$ (Clump A) and $\sim 150$ km s$^{-1}$ (B).
Although these clumps may not be collapsed,
the formal virial masses can be calculated from 
velocity dispersion, $\sigma_{v1}$, and radius, $r$, with
$3 \sigma_{v1}^2 r / G$, where $G$ is the gravitational constant.
We obtain
$\sim 1\times 10^{13} M_\odot$ (Clump A) and
$\sim 8\times 10^{12} M_\odot$ (B).
The three-dimensional density contrast
of LAEs, $\delta_n=\delta n/\bar{n}$,
for the average of these clumps is $\sim 80$,
which is comparable to
those of present-day clusters, $100-200$.
These clumps have the highest density
in our surveyed volume of
$9\times 10^5$ Mpc$^3$ in comoving units.
In the present-day Universe, this volume
typically contains two massive clusters with mass of
$1-3\times 10^{14} M_\odot$ \citep{reiprich2002}.
Thus, the discovered clumps are likely proto-clusters which are
ancestors of today's such clusters.

A particularly interesting feature is that these clumps
are very high concentrations of star-forming galaxies 
with a star-formation rate of $1-20 M_\odot$ yr$^{-1}$.
The star-formation rate density (SFRD) in these clumps
inferred from Ly$\alpha$ fluxes
is about 130 times higher than in the mean of
the whole 1 deg$^2$ field, mainly due to a large
overdensity, $\delta_n\sim 80$, of LAEs. 
This very high SFRD implies that
these clumps would be just producing
a number of galaxies in a short period  
by a burst of galaxy formation.
In contrast, present-day massive clusters are
dominated by old early-type galaxies, and show a deficit of
young star-forming galaxies \citep{dressler1980,butcher1984}.
Since the formation epochs of early-type galaxies
are estimated to be $z=2-5$ or earlier
(e.g. \citealt{kodama1998}),
galaxies residing in these clumps are likely
progenitors of the old early types seen in the core of
present-day clusters.
Therefore, we are probably witnessing forming clusters
where a number of present-day early-type galaxies are just
emerging as star-forming galaxies by a burst of galaxy formation.

\subsection{Implications for Galaxy Formation}
The Cold Dark Matter (CDM) model, the standard hierarchical
scenario, predicts that very small initial fluctuations
of mass density grow up
gradually with time to evolve into galaxies and LSSs.
However, the observed distribution of LAEs
appears much more inhomogeneous than the dark-matter
distribution,
since the amplitude of the matter fluctuations at $z=5.7$
is predicted to be only 1/5 of the current value.
If one wants to accept the CDM model,
a solution to this discrepancy is to assume that LAEs
are biased tracers of mass in a way to enhance
the mass fluctuations.
Previous observational studies of high-$z$ galaxies discuss the
clustering bias
\citep{adelberger1998,giavalisco2001,ouchi2001,shimasaku2004,ouchi2004b}.
Theoretically, such biasing can be produced,
if galaxies are formed at rare, highest peaks of yet small
fluctuations of matter density (e.g. \citealt{mo2002}).
From $\sigma_{20}$, we estimate that
the bias parameter, $b$,
the ratio of galaxy overdensity to matter overdensity,
is $b=3.4\pm 1.8$ on scales of LSSs (20 Mpc),
which is comparable to the prediction at $z\sim5$
from simulations of galaxy formation
based on the CDM model \citep{baugh1999,kauffmann1999}.
Thus, the CDM model may be consistent with the early
formation of LSSs that we discovered, though the model
needs to reproduce the characteristic shapes of observed
LSSs, long filaments and large voids.
On the other hand,
we estimate the $b$ value, in the CDM, which reproduces
the observed number of forming clusters (two in $9\times 10^5$ Mpc$^3$)
with the size and the overdensity, $\delta_n\sim 80$,
following the prescription of \citet{shimasaku2003}.
We obtain $b\sim 30$ for the two forming clusters,
which is about ten times
higher than the value for LSS scales, 
although this estimation of $b$ is based on small statistics and 
might include a large error due to cosmic variance.
This variety of $b$ could be due to
the dependence of $b$ on 
the clustering scale or galaxy density \citep{benson2001}. 
However, it is not yet clear how such a large difference in $b$
may be produced within the framework of the CDM model.
Our cosmic map may raise a question about
how and where galaxies are formed in the early Universe.

\acknowledgments
M. Ouchi acknowledges the support from the 
Hubble Fellowship program through grant HF-01176.01-A.


\clearpage

\clearpage

\clearpage

\clearpage

\clearpage

\begin{deluxetable}{ccclcc}
\tabletypesize{\scriptsize}
\tablecaption{Results of Spectroscopy
\label{tab:lae_spec}}
\tablewidth{0pt}
\tablehead{
\colhead{RA(J2000)} &
\colhead{Dec(J2000)} & 
\colhead{$i'_{\rm AB}$} &
\colhead{$z$} &
\colhead{$f$} &
\colhead{$SFR$}
}
\startdata
02:17:45.03 & -05:28:42.5 & 26.8 & 5.749\tablenotemark{a} & 2.0 & 5.9\\ 
02:20:13.36 & -04:51:09.3 & 26.7 & 5.742\tablenotemark{a} & 1.3 & 3.7\\
02:20:26.14 & -04:52:34.3 & 25.7 & 5.717 & 0.8 & 2.4\\
02:20:19.79 & -04:52:29.9 & 26.4 & 5.706 & 1.0 & 2.9\\ 
02:17:49.13 & -05:28:54.2 & 26.2 & 5.694\tablenotemark{a} & 5.0 & 14.5\\
02:17:50.00 & -05:27:08.2 & 27.4 & 5.693 & 0.7 & 1.9\\
02:17:47.32 & -05:26:48.0 & 25.5 & 5.692 & 0.7 & 2.1\\ 
02:17:48.48 & -05:31:27.0 & 26.6 & 5.687\tablenotemark{a} & 1.4 & 3.9\\
02:17:45.27 & -05:29:36.1 & 26.6 & 5.687\tablenotemark{a} & 4.1 & 11.7\\
02:17:43.35 & -05:28:07.1 & 26.2 & 5.683 & 7.0 & 20.0\\
02:18:23.82 & -05:32:05.3 & 27.1 & 5.678 & 1.0 & 2.8\\
02:18:19.13 & -05:33:11.7 & 27.1 & 5.674 & 2.0 & 5.7\\
02:18:20.88 & -05:33:21.9 & 26.5 & 5.670 & 0.2 & 0.7\\ 
02:18:14.43 & -05:32:49.1 & 26.5 & 5.670\tablenotemark{a} & 1.0 & 2.9\\
02:20:21.52 & -04:53:15.0 & 26.7 & 5.669 & 1.8 & 5.1\\
02:18:22.61 & -05:33:37.8 & 27.6 & 5.648 & 1.5 & 4.4\\
02:18:17.35 & -05:32:22.8 & 26.6 & 5.642\tablenotemark{a} & 2.9 & 8.2\\
02:17:43.31 & -05:31:35.0 & 26.5 & 5.624\tablenotemark{a} & 2.5 & 7.1\\
02:17:45.88 & -05:27:14.5 & 25.6 & 5.600 & 0.9 & 2.4\\
02:17:51.14 & -05:29:35.3 & 27.2 & 1.188\tablenotemark{b} & 1.7 & \nodata\\
02:20:12.15 & -04:49:50.7 & 27.4 & 1.179\tablenotemark{b} & 1.5 & \nodata\\
02:20:12.84 & -04:50:22.5 & 24.5 & 0.623\tablenotemark{c} & 11.9 & \nodata
\enddata
\tablecomments{
The quantities $i'_{\rm AB}$, $f$ and $SFR$ are 
the $i'$-band magnitude measured in a $2''$ diameter aperture,
the total line flux in units of 
$10^{-17}$ ergs $s^{-1}$ cm$^{-2}$,
and the star-formation rate in units of $M_\odot$ yr$^{-1}$, respectively.
The redshift, $z$, is measured from spectra summed over 
$0''.9$ around the source center, while the line flux, 
$f$, is obtained from total flux falling in the slit.
}
\tablenotetext{a}{Securely identified LAEs}
\tablenotetext{b}{[O{\sc ii}] emitters}
\tablenotetext{c}{[O{\sc iii}] emitter}
\end{deluxetable}


\begin{thebibliography}{}
\bibitem[Adelberger et al.(1998)]{adelberger1998} Adelberger, K.~L., 
  Steidel, C.~C., Giavalisco, M., Dickinson, M., Pettini, M., \& Kellogg, M.\ 
  1998, \apj, 505, 18 
\bibitem[Ajiki et al.(2003)]{ajiki2003} Ajiki, M., et al.\ 2003, 
  \aj, 126, 2091 
\bibitem[Baugh et al.(1999)]{baugh1999} Baugh, C.~M., Benson, 
A.~J., Cole, S., Frenk, C.~S., \& Lacey, C.~G.\ 1999, \mnras, 305, L21 
\bibitem[Benson et al.(2001)]{benson2001} Benson, A.~J., Frenk, 
C.~S., Baugh, C.~M., Cole, S., \& Lacey, C.~G.\ 2001, \mnras, 327, 1041 
\bibitem[Bertin \& Arnouts(1996)]{bertin1996} Bertin, E.\ \& Arnouts, S.\ 
  1996, \aaps, 117, 393
\bibitem[Bruzual \& Charlot(2003)]{bruzual2003} Bruzual, G.~\& 
  Charlot, S.\ 2003, \mnras, 344, 1000 
\bibitem[Butcher \& Oemler(1984)]{butcher1984} Butcher, H.~\& 
  Oemler, A.\ 1984, \apj, 285, 426 
\bibitem[Coleman, Wu, \& Weedman(1980)]{coleman1980} Coleman, G.~D., Wu, 
  C.-C., \& Weedman, D.~W.\ 1980, \apjs, 43, 393
\bibitem[Dressler(1980)]{dressler1980} Dressler, A.\ 1980, \apj, 
  236, 351 
\bibitem[Giavalisco \& Dickinson(2001)]{giavalisco2001} Giavalisco, 
  M.~\& Dickinson, M.\ 2001, \apj, 550, 177 
\bibitem[Gunn \& Stryker(1983)]{gunn1983} Gunn, J.~E.~\& Stryker, L.~L.\ 
  1983, \apjs, 52, 121
\bibitem[Hayashino et al.(2003)]{hayashino2003} Hayashino, T., et 
  al.\ 2003, Publications of the National Astronomical Observatory of Japan, 
  7, 33 
\bibitem[Hu et al.(2004)]{hu2004} Hu, E.~M., Cowie, L.~L., 
  Capak, P., McMahon, R.~G., Hayashino, T., \& Komiyama, Y.\ 2004, \aj, 127, 
  563 
\bibitem[Kashikawa et al.(2002)]{kashikawa2002} Kashikawa, N.~et al.\ 
  2002, \pasj, 54, 819
\bibitem[Kauffmann, Colberg, Diaferio, \& White(1999)]{kauffmann1999} 
  Kauffmann, G., Colberg, J.~M., Diaferio, A., \& White, S.~D.~M.\ 1999, 
  \mnras, 307, 529 
\bibitem[Kinney et al.(1996)]{kinney1996} Kinney, A.~L., Calzetti, D., 
  Bohlin, R.~C., McQuade, K., Storchi-Bergmann, T., \& Schmitt, H.~R.\ 1996, 
  \apj, 467, 38
\bibitem[Kodama, Arimoto, Barger, \& 
  Arag'on-Salamanca(1998)]{kodama1998} Kodama, T., Arimoto, N., 
  Barger, A.~J., \& Arag'on-Salamanca, A.\ 1998, \aap, 334, 99 
\bibitem[Kodama et al.(2004)]{kodama2004} Kodama, T., et al.\ 
  2004, \mnras, 350, 1005 
\bibitem[Madau(1995)]{madau1995} Madau, P.\ 1995, \apj, 441, 18
\bibitem[Miley et al.(2004)]{miley2004} Miley, G.~K., et al.\ 
  2004, \nat, 427, 47 
\bibitem[Miyazaki et al.(2002)]{miyazaki2002} Miyazaki, S.~et al.\ 
  2002, \pasj, 54, 833 
\bibitem[Mo \& White(2002)]{mo2002} Mo, H.~J.~\& White, 
  S.~D.~M.\ 2002, \mnras, 336, 112 
\bibitem[Rhoads et al.(2003)]{rhoads2003} Rhoads, J.~E., et al.\ 
  2003, \aj, 125, 1006 
\bibitem[Ouchi et al.(2001)]{ouchi2001} Ouchi, M.~et al.\ 2001, \apjl, 558, L83
\bibitem[Ouchi et al.(2003)]{ouchi2003} Ouchi, M.~et al.\ 2003, 
  \apj, 582, 60
\bibitem[Ouchi et al.(2004a)]{ouchi2004a} Ouchi, M., et al.\ 2004a, 
  \apj, 611, 660 
\bibitem[Ouchi et al.(2004b)]{ouchi2004b} Ouchi, M., et al.\ 2004b, 
  \apj, 611, 685 
\bibitem[Peebles (1980)]{peebles1980} Peebles, P. J. E. 1980, The 
  Large-Scale Structure of the Universe (Princeton: Princeton Univ. Press)
\bibitem[Reiprich \& B{\" o}hringer(2002)]{reiprich2002} Reiprich, 
  T.~H.~\& B{\" o}hringer, H.\ 2002, \apj, 567, 716 
\bibitem[Seaborne et al.(1999)]{seaborne1999} Seaborne, M.~D., et 
  al.\ 1999, \mnras, 309, 89 
\bibitem[Sekiguchi et al.(2004)]{sekiguchi2004} Sekiguchi, K.~et al.\ 2004, 
  Astrophysics and Space Science Library, 301, 169
\bibitem[Shimasaku et al.(2003)]{shimasaku2003} Shimasaku, K.~et al.\ 
  2003, \apjl, 586, L111
\bibitem[Shimasaku et al.(2004)]{shimasaku2004} Shimasaku, K., et 
  al.\ 2004, \apjl, 605, L93 
\bibitem[Steidel et al.(1998)]{steidel1998} Steidel, C.~C., 
  Adelberger, K.~L., Dickinson, M., Giavalisco, M., Pettini, M., \& Kellogg, 
  M.\ 1998, \apj, 492, 428 
\bibitem[Steidel et al.(2000)]{steidel2000} Steidel, C.~C., Adelberger, 
  K.~L., Shapley, A.~E., Pettini, M., Dickinson, M., \& Giavalisco, M.\ 2000, 
  \apj, 532, 170
\bibitem[Venemans et al.(2002)]{venemans2002} Venemans, B.~P.~et 
  al.\ 2002, \apjl, 569, L11 
\bibitem[Venemans et al.(2004)]{venemans2004} Venemans, B.~P., et 
al.\ 2004, \aap, 424, L17 
\bibitem[Yagi et al.(2002)]{yagi2002} Yagi, M., Kashikawa, N., 
  Sekiguchi, M., Doi, M., Yasuda, N., Shimasaku, K., \& Okamura, S.\ 2002, 
  \aj, 123, 66
\end{thebibliography}
\end{document}